\input amstex
\documentstyle{conm-p}
\NoBlackBoxes

\issueinfo{00}
  {}
  {}
  {XXXX}

\topmatter
\title The Borg-Marchenko Theorem with a Continuous Spectrum\endtitle
\author Tuncay Aktosun and Ricardo Weder
\endauthor
\leftheadtext{TUNCAY AKTOSUN AND RICARDO WEDER}%

\address Department of Mathematics, University of Texas at Arlington,
Arlington,  TX 76019, USA \endaddress

\email aktosun\@uta.edu\endemail

\thanks
The research leading to this
article was supported in part by the National Science Foundation
under grant DMS-0204437, the Department of Energy under grant
DE-FG02-01ER45951, Universidad Nacional Aut\'onoma de M\'exico
under Proyecto PAPIIT-DGAPA IN 101902, and by
CONACYT under Proyecto P42553-F.
\endthanks

\address Instituto de Investigaciones en
Matem\'aticas Aplicadas y en Sistemas, Universidad Nacional Aut\'onoma de M\'exico,
Apartado Postal 20-726, IIMAS-UNAM, M\'exico DF 01000, M\'exico \endaddress

\email weder\@servidor.unam.mx\endemail

\thanks The second author is Fellow Sistema Nacional de Investigadores.
\endthanks




\abstract
 The Schr\"odinger equation is considered on the half
line with a selfadjoint boundary condition
when the potential is real valued, integrable, and has
a finite
first moment. It is proved that the potential and the two boundary
conditions are uniquely determined by a set of spectral
data containing the
discrete eigenvalues for a boundary condition at the origin, the
continuous part of the spectral measure for that boundary
condition, and a subset of the discrete eigenvalues for a
different boundary condition. This result provides a generalization
of the celebrated uniqueness
theorem of Borg and Marchenko using two sets of discrete
spectra to the case where there is also a continuous spectrum.
The proof employed yields a method to recover the potential and the two
boundary conditions, and it also constructs data sets used in
various inversion methods. A comparison is
made with the uniqueness result of Gesztesy and Simon using Krein's
spectral shift function as the inversion data.

\endabstract

\endtopmatter

\document

\head 1. Introduction\endhead

Consider the Schr\"odinger equation on the half line
$$-\psi''+V(x)\,\psi=k^2\psi,\qquad x\in\bold R^+,\tag 1.1$$
where the prime denotes the derivative with respect to $x,$
the potential $V$ is real valued and
measurable, and $\int_0^\infty dx\,(1+x)\,|V(x)|$ is finite.
Such potentials are said to make up the Faddeev class.
Let $H_\alpha$ for a fixed $\alpha\in(0,\pi]$ denote
the unique selfadjoint realization [1] of the
corresponding Schr\"odinger operator on $L^2(\bold R^+)$
with the boundary condition
$$\sin\alpha\cdot
\psi'(k,0)+\cos\alpha\cdot\psi(k,0)=0,\tag 1.2$$
which can also be written as
$$\cases \psi'(k,0)+\cot\alpha\cdot\psi(k,0)=0,\qquad \alpha\in(0,\pi),\\
\psi(k,0)=0,\qquad \alpha=\pi.\endcases$$
Note that $\alpha\mapsto\cot\alpha$ is a monotone decreasing mapping
of $(0,\pi)$ onto $\bold R,$ and hence
$\alpha$ is uniquely determined by $\cot\alpha.$

It is known [1,2] that $H_\alpha$ has no
positive or zero eigenvalues, has no singular-continuous
spectrum, has at most a finite number of (simple) negative
eigenvalues,
and its absolutely-continuous spectrum consists of
$k^2\in[0,+\infty).$

Borg [3] and Marchenko [4] independently analyzed (1.1)
with the boundary condition (1.2) when there is no
continuous spectrum.
They showed [3-5] that two sets of discrete spectra
associated with two distinct boundary conditions at $x=0$ (with a
fixed boundary condition, if any, at $x=+\infty$) uniquely
determine the potential and the two boundary conditions.

A continuous spectrum often appears in applications, and
it usually arises when the potential
vanishes at infinity.
In our paper we present an extension of the
celebrated Borg-Marchenko result in the presence of
a continuous spectrum; namely, we show
that the potential and the two boundary conditions are uniquely determined
by an appropriate data set containing the discrete eigenvalues and
the continuous part of the spectral measure corresponding to one
boundary condition and a subset of the discrete
eigenvalues corresponding to a different boundary condition.

Another extension
of the Borg-Marchenko result with a continuous
spectrum is given by Gesztesy and Simon [6], where a
uniqueness result is presented when the corresponding
Krein's spectral shift
function
is used as the data
in the class of real-valued potentials that are
integrable on $[0,R]$ for all $R>0.$
In our generalization of the Borg-Marchenko
theorem, we specify the data in terms of a subset
of the spectral measure; namely, the amplitude of the Jost
function and the eigenvalues.
The connection between the data used in [6] and ours
is analyzed in Section~5.

The problem under study has
applications in the acoustical analysis of the human vocal tract.
The related inverse problem can be described [7-9] as
determining a scaled curvature
of the duct of the vocal tract when a constant-frequency sound is
uttered. Such
an inverse problem has important applications in speech recognition.

Our paper is organized as follows. In Section~2 we introduce
the preliminary material related to the Jost function, the phase shift,
and the spectral measure. In Section~3 we present our generalized
Borg-Marchenko theorem. In Section~4 we briefly outline
the Gel'fand-Levitan, Marchenko, and the Faddeev-Marchenko
procedures to recover
the potential.
In Section~5 we show how
the data
used in our theorem uniquely constructs Krein's spectral shift function
and vice versa. Finally, in Section~6 we present two examples
to illustrate the theory presented in our paper.

\head 2. Preliminaries \endhead

Recall that the Jost function associated with
$H_\alpha$ is defined as
[2,10-12]
$$F_\alpha(k):=\cases -i[f'(k,0)+\cot\alpha\cdot f(k,0)],\qquad
\alpha\in(0,\pi),\\
f(k,0),\qquad \alpha=\pi,\endcases\tag 2.1$$
with $f(k,x)$
denoting the Jost solution to (1.1) satisfying the asymptotics
$$f(k,x)=e^{ikx}[1+o(1)],\quad
f'(k,x)=ik e^{ikx}[1+o(1)],\qquad x\to+\infty.\tag 2.2$$
 From (1.1) and (2.2) it follows that
$$f(-k,0)=f(k,0)^\ast,\quad
f'(-k,0)=f'(k,0)^\ast,\qquad k\in\bold R,\tag 2.3$$
where the asterisk denotes complex conjugation, and hence
for $k\in\bold R$ we get
$$F_\alpha(-k)=\cases -F_\alpha(k)^\ast,\qquad
\alpha\in(0,\pi),\\
F_\pi(k)^\ast,\qquad
\alpha=\pi.\endcases\tag 2.4$$
We use $\bold C^+$ for the upper half complex plane,
$\overline{\bold C^+}:=\bold C^+\cup\bold R$ for its closure, and
$\bold I^+:=i(0,+\infty)$ for the positive imaginary axis in $\bold C^+.$

It is known [2,10-12] that $F_\alpha(k)$ is analytic
in $\bold C^+$ and continuous in $\overline{\bold C^+},$ the zeros in $\bold C^+$ of
$F_\alpha,$
if any, can only occur on $\bold I^+$ and such zeros are all simple,
$F_\alpha(k)\ne 0$ for
$k\in\bold R\setminus\{0\},$ and
that either $F_\alpha(k)$ is nonzero at $k=0$ (generic case)
or it has a simple zero there (exceptional case). Because of (2.4),
knowledge of $F_\alpha(k)$ for $k\in\bold R^+$ is equivalent to that
for $k\in\bold R.$
Let $k=i\kappa_{\alpha j}$ for $j=1,\dots,N_\alpha$
represent the zeros of
$F_\alpha$ on $\bold I^+.$
Thus,
the set $\{-\kappa_{\alpha j}^2\}_{j=1}^{N_\alpha}$
corresponds to the discrete eigenvalues of $H_\alpha.$

The negative of the phase of the Jost function $F_\alpha$ is usually known
as the phase shift $\phi_\alpha,$ i.e.
$$e^{-i\phi_\alpha(k)}:=\displaystyle\frac{F_\alpha(k)}{|F_\alpha(k)|},\qquad k\in\bold R,\tag 2.5$$
where it is understood that
$\phi_\alpha(+\infty)=0.$
For $k\in\bold R,$ the phase shift $\phi_\alpha$ satisfies
$$\phi_\alpha(-k)=\cases \pi-\phi_\alpha(k),\qquad \alpha\in(0,\pi),\\
-\phi_\pi(k),\qquad \alpha=\pi.\endcases\tag 2.6$$
The scattering matrix $S_\alpha(k)$ for $k\in\bold R$
associated with $H_\alpha$ is defined as
$$S_\alpha(k):=e^{2i\phi_\alpha(k)}=
\cases -\displaystyle\frac{F_\alpha(-k)}{F_\alpha(k)},
\qquad \alpha\in(0,\pi),\\
\displaystyle\frac{F_\pi(-k)}{F_\pi(k)},\qquad \alpha=\pi.
\endcases\tag 2.7$$

The number of discrete eigenvalues $N_\alpha$ is related to
the phase shift $\phi_\alpha$ by Levinson's theorem [2,11], i.e.
$$\phi_\alpha(0^+)=\cases \left(N_\alpha+\displaystyle\frac{1+d_\alpha}{2}\right)\pi,
\qquad \alpha\in(0,\pi),\\
\left(N_\pi+\displaystyle\frac{d_\pi}{2}\right)\pi,
\qquad \alpha=\pi,\endcases
\tag 2.8$$
where we have defined
$$d_\alpha:=\cases 0,\qquad \text { if } F_\alpha(0)\ne 0,\\
1,\qquad \text { if } F_\alpha(0)=0.\endcases$$

The spectral measure $\rho_\alpha$
corresponding to $H_\alpha$ can be determined via [2,10-12]
$$d\rho_\alpha(\lambda)=\cases
\displaystyle\frac{\sqrt{\lambda}}{\pi}\,\displaystyle\frac{1}
{|F_\alpha(\sqrt{\lambda})|^2}\,d\lambda,\qquad
\lambda >0,\\
\displaystyle\sum_{j=1}^{N_\alpha} g_{\alpha j}^2\,
\delta(\lambda+\kappa_{\alpha j}^2)\,d\lambda, \qquad \lambda <0,
\endcases$$
where $\delta(\cdot)$ is the Dirac
delta distribution, $\lambda:=k^2,$ and the norming constants $g_{\alpha j}$ are given
by
$$g_{\alpha j}:=\cases \displaystyle\frac{|f(i\kappa_{\alpha j},0)|}
{||f(i\kappa_{\alpha j},\cdot)||},\qquad \alpha\in(0,\pi),\\
\displaystyle\frac{|f'(i\kappa_{\pi j},0)|}{||f(i\kappa_{\pi
j},\cdot)||},\qquad \alpha=\pi,\endcases\tag 2.9$$
with
$||\cdot||$ denoting the standard norm in $L^2(0,+\infty).$
It is known [2,10-12] that $\{V,\alpha\}$ is
uniquely determined by the corresponding spectral measure $\rho_\alpha$ and that the
reconstruction can be achieved by solving the
Gel'fand-Levitan integral equation.

\head 3. The Borg-Marchenko Theorem with a Continuous Spectrum \endhead

Our generalized Borg-Marchenko theorem consists of identifying the appropriate
data set leading to the unique determination of the potential $V$
in (1.1) and the two distinct boundary parameters $\alpha$ and $\beta$
in (1.2) with $0<\beta<\alpha\le \pi.$
Here, we briefly state our theorem and summarize the steps to
construct $\{V,\beta,\alpha\}$ and refer the reader to [12] for the
proof and further details.

Let the set $\{-\kappa_{\beta j}^2\}_{j=1}^{N_\beta}$
correspond to the discrete eigenvalues of $H_\beta.$
We use $F_\beta$ to denote the Jost function associated with
$H_\beta,$ and it is obtained by replacing $\alpha$ with $\beta$
on the right hand side of (2.1).

Our motivation is
as follows. Assume that we are given some data set
$\Cal D,$ which contains $|F_\alpha(k)|$ for $k\in\bold R,$ the whole
set $\{\kappa_{\alpha j}\}_{j=1}^{N_\alpha},$ and a subset of
$\{\kappa_{\beta j}\}_{j=1}^{N_\beta}$ consisting of $N_\alpha$
elements. Alternatively, our data $\Cal D$ may include
$|F_\beta(k)|$ for $k\in\bold R$ and the sets $\{\kappa_{\alpha
j}\}_{j=1}^{N_\alpha}$ and $\{\kappa_{\beta j}\}_{j=1}^{N_\beta}.$
Does $\Cal D$ uniquely determine $\{V,\alpha,\beta\}$?
If not, what additional
information do we need to include in
$\Cal D$  for the unique determination?
Can we also present a constructive method to
recover $\{V,\alpha,\beta\}$ from $\Cal D$ or from a data set obtained
by some augmentation of $\Cal D$?

Since $0<\beta<\alpha\le \pi,$ from the interlacing properties
of eigenvalues, it is known [2,10-12] that either
$N_\beta=N_\alpha,$ in which case we have
$$0<\kappa_{\alpha 1}<\kappa_{\beta 1}
<\kappa_{\alpha 2}<\kappa_{\beta 2}<\dots< \kappa_{\alpha
N_\alpha}<\kappa_{\beta N_\alpha},\tag 3.1$$
or else we have $N_\beta=N_\alpha+1,$ in which case we get
$$0<\kappa_{\beta 1}<\kappa_{\alpha 1}<
\kappa_{\beta 2}<\kappa_{\alpha 2}<\dots< \kappa_{\alpha
N_\alpha}<\kappa_{\beta N_\beta}.\tag 3.2$$

There are eight distinct cases to consider
depending on whether $\alpha\in(0,\pi)$ or $\alpha=\pi,$
whether $N_\beta=N_\alpha$ or $N_\beta=N_\alpha+1,$
and whether the data set used contains $|F_\alpha|$ or
$|F_\beta|.$ So, let us define
the data sets $\Cal D_1,\dots,\Cal D_8$ as follows [12]:
$$\Cal D_1:=\{h_{\beta\alpha},|F_\alpha(k)| \text{ for } k\in\bold R,
\{\kappa_{\alpha j}\}_{j=1}^{N_\alpha},
\{\kappa_{\beta j}\}_{j=1}^{N_\beta}\},\tag 3.3$$
$$\Cal D_2:=\{\beta,|F_\pi(k)| \text{ for } k\in\bold R,
\{\kappa_{\pi j}\}_{j=1}^{N_\pi},
\{\kappa_{\beta j}\}_{j=1}^{N_\beta}\},\tag 3.4$$
$$\Cal D_3:=\{h_{\beta\alpha},|F_\alpha(k)| \text{ for } k\in\bold R,
\{\kappa_{\alpha j}\}_{j=1}^{N_\alpha}, {\text{$N_\alpha$-element subset of}}
\ \{\kappa_{\beta j}\}_{j=1}^{N_\beta}\},\tag 3.5$$
$$\Cal D_4:=\{\beta,|F_\pi(k)| \text{ for } k\in\bold R,
\{\kappa_{\pi j}\}_{j=1}^{N_\pi}, {\text{$N_\pi$-element subset of}}
\ \{\kappa_{\beta j}\}_{j=1}^{N_\beta}\},\tag 3.6$$
$$\Cal D_5:=\{h_{\beta\alpha},|F_\beta(k)| \text{ for } k\in\bold R,
\{\kappa_{\alpha j}\}_{j=1}^{N_\alpha}, \{\kappa_{\beta
j}\}_{j=1}^{N_\beta}\},\tag 3.7$$
$$\Cal D_6:=\{|F_\beta(k)| \text{ for } k\in\bold R,
\{\kappa_{\pi j}\}_{j=1}^{N_\pi}, \{\kappa_{\beta
j}\}_{j=1}^{N_\beta}\},\tag 3.8$$
$$\Cal D_7:=\{\beta,h_{\beta\alpha},|F_\beta(k)| \text{ for } k\in\bold R,
\{\kappa_{\alpha j}\}_{j=1}^{N_\alpha}, \{\kappa_{\beta
j}\}_{j=1}^{N_\beta}\},\tag 3.9$$
$$\Cal D_8:=\{\beta,|F_\beta(k)| \text{ for } k\in\bold R,
\{\kappa_{\pi j}\}_{j=1}^{N_\pi}, \{\kappa_{\beta
j}\}_{j=1}^{N_\beta}\},\tag 3.10$$
where we let
$$h_{\beta\alpha}:=\cot\beta-\cot\alpha.\tag 3.11$$
Note that $h_{\beta\alpha}>0$ when
$0<\beta<\alpha<\pi.$
The data sets $\Cal D_1,\Cal D_3,\Cal D_5,\Cal D_7$ correspond to
$\alpha\in(0,\pi)$ and the sets $\Cal D_2,\Cal D_4,\Cal D_6,\Cal D_8$ to
$\alpha= \pi;$ the sets $\Cal D_1,\Cal D_2,\Cal D_3,\Cal D_4$ contain
$|F_\alpha|$ whereas the sets $\Cal D_5,\Cal D_6,\Cal D_7,\Cal D_8$ contain
$|F_\beta|;$ the sets $\Cal D_1,\Cal D_2,\Cal D_5,\Cal D_6$ correspond to
$N_\beta=N_\alpha$ whereas the sets $\Cal D_3,\Cal D_4,\Cal D_7,\Cal D_8$
to $N_\beta=N_\alpha+1.$ Our generalized Borg-Marchenko theorem can be stated as follows.

\proclaim {Theorem 3.1} Let the realizations $H_\alpha$ and $H_\beta$
for some $0<\beta<\alpha\le\pi$
correspond to a potential $V$ in the Faddeev class
with the boundary conditions
identified by $\alpha$ and $\beta,$ respectively. Then, each of the data
sets $\Cal D_j$ with $j=1,\dots,8$ uniquely determines the corresponding
$\{V,\alpha,\beta\}.$
\endproclaim

The proof
of the above theorem
provides a method to recover $\alpha$ and $\beta$ as well as $F_\alpha(k)$
and $F_\beta(k)$ for $k\in\overline{\bold C^+},$ thus also allowing us to construct
the data sets used as input in various inversion methods to determine $V.$
For the proof and details we refer the reader to [12], and here we only briefly
outline the steps involved.
Using $\Cal D_j$ for each of $j=1,2,5,6,7,8$
we first construct $\text{Re}[\Lambda_j(k)]$ for $\bold R,$ where
$$\text{Re}[\Lambda_1(k)]=\displaystyle\frac{k\,h_{\beta\alpha}}{|F_\alpha(k)|^2}
\displaystyle\prod_{j=1}^{N_\alpha}\displaystyle\frac{k^2+\kappa_{\alpha j}^2}{k^2+\kappa_{\beta j}^2},
\quad\text{Re}[\Lambda_2(k)]=-1+\displaystyle\frac{1}{|F_\pi(k)|^2}
\displaystyle\prod_{j=1}^{N_\pi}\displaystyle\frac{k^2+\kappa_{\pi j}^2}
{k^2+\kappa_{\beta j}^2},$$
$$\text{Re}[\Lambda_3(k)]=\displaystyle\frac{h_{\beta\alpha}\,k^2}{|F_\alpha(k)|^2}
\displaystyle\frac{\displaystyle\prod_{j=1}^{N_\alpha}(k^2+\kappa_{\alpha j}^2)}
{\displaystyle\prod_{j=1}^{N_\beta}(k^2+\kappa_{\beta j}^2)},
\quad\text{Re}[\Lambda_4(k)]=-1+\displaystyle\frac{k^2}{|F_\pi(k)|^2}
\displaystyle\frac{\displaystyle\prod_{j=1}^{N_\pi}(k^2+\kappa_{\pi j}^2)}
{\displaystyle\prod_{j=1}^{N_\beta}(k^2+\kappa_{\beta j}^2)},
$$
$$\text{Re}[\Lambda_5(k)]=-h_{\beta\alpha}+\displaystyle\frac{k^2\,h_{\beta\alpha}}{|F_\beta(k)|^2}
\displaystyle\prod_{j=1}^{N_\beta}\displaystyle\frac{k^2+\kappa_{\beta
j}^2}{k^2+\kappa_{\alpha j}^2},
$$
$$\text{Re}[\Lambda_6(k)]=-1+\displaystyle\frac{k^2}{|F_\beta(k)|^2}
\displaystyle\prod_{j=1}^{N_\beta}\displaystyle\frac{k^2+\kappa_{\beta
j}^2}{k^2+\kappa_{\pi j}^2},$$
$$\text{Re}[\Lambda_7(k)]=
h_{\beta\alpha}-\displaystyle\frac{h_{\beta\alpha}}{|F_\beta(k)|^2}
\displaystyle\frac{\displaystyle\prod_{j=1}^{N_\beta}(k^2+\kappa_{\beta j}^2)}
{\displaystyle\prod_{j=1}^{N_\beta-1}(k^2+\kappa_{\alpha j}^2)},
$$
$$\text{Re}[\Lambda_8(k)]=-1+\displaystyle\frac{1}{|F_\beta(k)|^2}
\displaystyle\frac{\displaystyle\prod_{j=1}^{N_\beta}(k^2+\kappa_{\beta j}^2)}
{\displaystyle\prod_{j=1}^{N_\beta-1}(k^2+\kappa_{\alpha j}^2)}.$$
Then, using the Schwarz integral formula
$$\Lambda_j(k)=\displaystyle\frac{1}{\pi i}\int_{-\infty}^\infty
\displaystyle\frac{dt}{t-k-i0^+}\,
\text{Re}[\Lambda_j(t)],\qquad k\in\overline{\bold C^+},$$
we uniquely construct $\Lambda_j,$
where $0^+$ in the integrand indicates that
the values of $\Lambda_j(k)$
for $k\in\bold R$ must be obtained via a limit from
$\bold C^+.$
We get
$$\Lambda_1(k)=-i+i\,\displaystyle\frac{F_\beta(k)}{F_\alpha(k)}
\displaystyle\prod_{j=1}^{N_\alpha}\displaystyle\frac{k^2+\kappa_{\alpha j}^2}{k^2+\kappa_{\beta j}^2},$$
$$\Lambda_2(k)=-1-\displaystyle\frac{1}{k}\displaystyle\frac{F_\beta(0)}{F_\pi(0)}
\displaystyle\prod_{j=1}^{N_\pi}\displaystyle\frac{\kappa_{\pi j}^2}{\kappa_{\beta j}^2}
+\displaystyle\frac{1}{k}\displaystyle\frac{F_\beta(k)}{F_\pi(k)}
\displaystyle\prod_{j=1}^{N_\pi}\displaystyle\frac{k^2+\kappa_{\pi j}^2}
{k^2+\kappa_{\beta j}^2},$$
$$\Lambda_3(k)=ik\,\displaystyle\frac{F_\beta(k)}{F_\alpha(k)}
\displaystyle\frac{\displaystyle\prod_{j=1}^{N_\alpha}(k^2+\kappa_{\alpha j}^2)}
{\displaystyle\prod_{j=1}^{N_\beta}(k^2+\kappa_{\beta j}^2)},\quad
\Lambda_4(k)=-1+k\,\displaystyle\frac{F_\beta(k)}{F_\pi(k)}
\displaystyle\frac{\displaystyle\prod_{j=1}^{N_\pi}(k^2+\kappa_{\pi j}^2)}
{\displaystyle\prod_{j=1}^{N_\beta}(k^2+\kappa_{\beta j}^2)},$$
$$\Lambda_5(k)=ik-h_{\beta\alpha}-
ik\,\displaystyle\frac{F_\alpha(k)}{F_\beta(k)}
\displaystyle\prod_{j=1}^{N_\beta}\displaystyle\frac{k^2+\kappa_{\beta
j}^2}{k^2+\kappa_{\alpha j}^2},
\quad\Lambda_6(k)=-1+k\,
\displaystyle\frac{F_\pi(k)}{F_\beta(k)}
\displaystyle\prod_{j=1}^{N_\beta}\displaystyle\frac{k^2+\kappa_{\beta
j}^2}{k^2+\kappa_{\pi j}^2},$$
$$\Lambda_7(k)=-ik+h_{\beta\alpha}-
\displaystyle\frac{i}{k}\,\displaystyle\frac{F_\alpha(0)}{F_\beta(0)}
\displaystyle\frac{\displaystyle\prod_{j=1}^{N_\beta}\kappa_{\beta j}^2}
{\displaystyle\prod_{j=1}^{N_\beta-1}\kappa_{\alpha j}^2}+
\displaystyle\frac{i}{k}\,\displaystyle\frac{F_\alpha(k)}{F_\beta(k)}
\displaystyle\frac{\displaystyle\prod_{j=1}^{N_\beta}(k^2+\kappa_{\beta j}^2)}
{\displaystyle\prod_{j=1}^{N_\beta-1}(k^2+\kappa_{\alpha j}^2)},$$
$$\Lambda_8(k)=-1-
\displaystyle\frac{1}{k}\,\displaystyle\frac{F_\pi(0)}{F_\beta(0)}
\displaystyle\frac{\displaystyle\prod_{j=1}^{N_\beta}\kappa_{\beta j}^2}
{\displaystyle\prod_{j=1}^{N_\beta-1}\kappa_{\alpha j}^2}+
\displaystyle\frac{1}{k}\,\displaystyle\frac{F_\pi(k)}{F_\beta(k)}
\displaystyle\frac{\displaystyle\prod_{j=1}^{N_\beta}(k^2+\kappa_{\beta j}^2)}
{\displaystyle\prod_{j=1}^{N_\beta-1}(k^2+\kappa_{\alpha j}^2)}.$$
Each $\Lambda_j(k)$
is analytic in $\bold C^+,$ continuous in $\overline{\bold C^+},$ and $O(1/k)$ as
$k\to\infty$ in $\overline{\bold C^+}.$

Next, with the help of $\Cal D_j$ and
$\Lambda_j,$ we uniquely construct
$\alpha,$ $\beta,$ $F_\alpha,$ and $F_\beta$
by using the procedure given in Section~4 of [12].
Having these four quantities, we can construct
$V$ by using one of the available inversion methods [2,10-12],
three of which are outlined in Section~4.
Note that $f(k,0)$ and $f'(k,0)$ can then also be constructed via
$$f(k,0)=\cases\displaystyle\frac{i}{h_{\beta\alpha}}\left[F_\beta(k)-F_\alpha(k)\right],
\qquad \alpha,\beta\in(0,\pi),\\
F_\pi(k),\endcases$$
$$f'(k,0)=\cases \displaystyle\frac{i}{h_{\beta\alpha}}\left[\cot\beta\cdot
F_\alpha(k)-\cot\alpha\cdot F_\beta(k)\right],\qquad \alpha,\beta\in(0,\pi),\\
i\,F_\beta(k)-\cot\beta\cdot F_\pi(k),\qquad\beta\in(0,\pi).\endcases$$

The constructions from $\Cal D_j$ with $j=3,4$ involve an extra step because
in each of those two cases exactly one of the discrete eigenvalues needed to construct
$\text{Re}[\Lambda_j(k)]$ for $k\in\bold R$ is not contained in $\Cal D_j.$
As a result, we first construct a one-parameter family of
$\Lambda_j(k)$ for $k\in\overline{\bold C^+}$ and then determine the eigenvalue
missing in $\Cal D_j$ by taking a nontangential limit as $k\to\infty$
on $\bold C^+.$ Once the missing discrete eigenvalue is at hand, the
corresponding $\Lambda_j(k)$ for $k\in\overline{\bold C^+}$
is uniquely determined as well. We can then proceed as
in the case with $j=1,2,5,6,7,8$ in order
to determine $\alpha$ and $\beta,$ $F_\alpha(k)$ and $F_\beta(k)$
for $k\in\overline{\bold C^+},$ the potential $V,$ and any other relevant
quantities.

\head 4. Reconstruction of the Potential \endhead

In Section~3 we have outlined the construction of the quantities
$\alpha,$ $\beta,$ and $F_\alpha(k)$ and $F_\beta(k)$
for $k\in\overline{\bold C^+}$ by using each of the data
sets $\Cal D_j$ with $j=1,\dots,8$ given in (3.3)-(3.10), respectively.
In this section we outline three methods to
briefly illustrate the use of
such quantities to recover the potential $V.$

In the Gel'fand-Levitan method [2,10-12], one forms the input data
$\Cal G_\alpha$
given by
$$\Cal G_\alpha:=\{|F_\alpha(k)|
\text{ for } k\in\bold R,\{\kappa_{\alpha j}\}_{j=1}^{N_\alpha},
\{g_{\alpha j}\}_{j=1}^{N_\alpha}\},$$
where the $g_{\alpha j}$ are the norming constants appearing in (2.9)
and they can also be obtained from (cf. (3.25) of [12])
$$g_{\alpha j}=\cases
\displaystyle\sqrt{\displaystyle\frac{2i\kappa_{\alpha j}\,F_\beta(i\kappa_{\alpha j})}
{h_{\beta\alpha}\,\dot
F_\alpha(i\kappa_{\alpha j})}},\qquad 0<\beta<\alpha<\pi,\\
\displaystyle\sqrt{\displaystyle\frac{2\,\kappa_{\pi j}\,F_\beta(i\kappa_{\pi j})}{\dot
F_\pi(i\kappa_{\pi j})}},\qquad 0<\beta<\alpha=\pi,\endcases$$
with an overdot indicating the $k$-derivative.
The
corresponding potential $V$ is uniquely recovered via
$$V(x)=2\,\displaystyle\frac{d}{dx}A_\alpha(x,x^-),$$
where $A_\alpha(x,y)$ is obtained by solving
the Gel'fand-Levitan integral equation
$$A_\alpha(x,y)+G_\alpha(x,y)+\int_0^x dz\,
G_\alpha(y,z)\,A_\alpha(x,z)=0,\qquad 0\le y<x,$$
with the kernel $G_\alpha(x,y)$ for $\alpha\in(0,\pi)$ given by
$$\aligned
G_\alpha(x,y):=\displaystyle\frac{1}{\pi}\int_{-\infty}^\infty
& dk\,\left[\displaystyle\frac{k^2}{|F_\alpha(k)|^2}-1\right]\left(\cos
kx\right) \left(\cos ky\right)\\
&+ \displaystyle\sum_{j=1}^{N_{\alpha}}
g_{\alpha j}^2 \left(\cosh
\kappa_{\alpha j}x\right)\left(\cosh \kappa_{\alpha j}y\right),\endaligned$$
and the kernel $G_\pi(x,y)$ given by
$$\aligned
G_\pi(x,y):=
\displaystyle\frac{1}{\pi}\int_{-\infty}^\infty
& dk\,\left[\displaystyle\frac{1}{|F_\pi(k)|^2}-1\right]\left(\sin
kx\right) \left(\sin ky\right)\\
& + \displaystyle\sum_{j=1}^{N_{\pi}}
\displaystyle\frac{g_{\pi j}^2}{\kappa_{\pi j}^2} \left(\sinh
\kappa_{\pi j}x\right)\left(\sinh \kappa_{\pi j}y\right).\endaligned$$

In the Marchenko method [2,11,12]
one forms the input data $\Cal M_\alpha$
given by
$$\Cal M_\alpha:=\{S_\alpha(k) \text{ for } k\in\bold R,\{\kappa_{\alpha j}\}_
{j=1}^{N_\alpha},\{m_{\alpha j}\}_{j=1}^{N_\alpha}\},$$
where $S_\alpha$ is the scattering matrix defined in (2.7) and the norming
constants $m_{\alpha j}$ can be obtained
by using
$$m_{\alpha j}:=\displaystyle\frac{1}{||f(i\kappa_{\alpha j},\cdot)||},\qquad
j=1,\dots,N_\alpha,$$
or equivalently via (cf. (3.26) of [12])
$$m_{\alpha j}=\cases
\displaystyle\sqrt{\displaystyle\frac{-2i\kappa_{\alpha j}\,h_{\beta\alpha}}
{F_\beta(i\kappa_{\alpha j})\,\dot
F_\alpha(i\kappa_{\alpha j})}},\qquad 0<\beta<\alpha<\pi,\\
\displaystyle\sqrt{\displaystyle\frac{-2\,\kappa_{\pi j}}
{F_\beta(i\kappa_{\pi j})\,\dot
F_\pi(i\kappa_{\pi j})}},\qquad 0<\beta<\alpha=\pi.\endcases$$
The potential $V$ is uniquely recovered as
$$V(x)=-2\,\displaystyle\frac{d}{dx}K(x,x^+),$$
where $K(x,y)$ is obtained by solving
the Marchenko integral equation
$$K(x,y)+M_\alpha(x+y)+\int_x^\infty dz\,M_\alpha(y+z)\,K(x,z)=0,\qquad
0<x<y,$$
with the kernel
$$M_\alpha(y):=\cases\displaystyle\frac{1}{2\pi}
\int_{-\infty}^\infty dk\,[S_\alpha(k)-1]\,e^{iky}
+\displaystyle\sum_{j=1}^{N_{\alpha}}m_{\alpha j}^2\,e^{-\kappa_{\alpha j}y},
\qquad \alpha\in(0,\pi),\\
\displaystyle\frac{1}{2\pi}
\int_{-\infty}^\infty dk\,[1-S_\pi(k)]\,e^{iky}
+\displaystyle\sum_{j=1}^{N_{\pi}}m_{\pi j}^2\,e^{-\kappa_{\pi j}y},
\qquad \alpha=\pi.\endcases$$

In the right Faddeev-Marchenko method [2,11,12], by viewing the potential
$V$ on the full line and setting
$V\equiv 0$ for $x<0,$
one forms the input data $\Cal F_{\text r}$
given by
$$\Cal F_{\text r}:=\{L(k) \text{ for } k\in\bold R,\{\tau_j\}_{j=1}^N,
\{c_{\text{r}j}\}_{j=1}^N\},$$
where $L$ is the left reflection coefficient
given by
$$L(k)=\displaystyle\frac{ik\,f(k,0)-f'(k,0)}{ik\,f(k,0)+f'(k,0)},
\qquad k\in\overline{\bold C^+},$$
or equivalently
$$L(k)=\cases
\displaystyle\frac{(k-i\cot\beta)\,F_\alpha(k)-
(k-i\cot\alpha)\,F_\beta(k)}{(k+i\cot\beta)\,F_\alpha(k)-
(k+i\cot\alpha)\,F_\beta(k)},\qquad \alpha,\beta\in(0,\pi),\\
\displaystyle\frac{(k-i\cot\beta)\,F_\pi(k)-
F_\beta(k)}{(k+i\cot\beta)\,F_\pi(k)+
F_\beta(k)},\qquad \beta\in(0,\pi).\endcases$$
The $N$ positive constants $\tau_j$ correspond to
the (simple) poles of $L(k)$ on $\bold I^+,$ and the
$c_{\text{r}j}$ are the norming constants
that can be obtained via
$$c_{\text{r}j}=\displaystyle\sqrt{-i\,\text{Res}(L,i\tau_j)},\qquad
j=1,\dots,N.$$
The potential $V$
can be uniquely reconstructed as
$$V(x)=2\,\displaystyle\frac{dB_{\text{r}}(x,0^+)}{dx},$$
where $B_{\text{r}}(x,y)$ is the solution to the right
Faddeev-Marchenko integral equation
$$B_{\text{r}}(x,y)+\Omega_{\text{r}}(-2x+y)+\int_0^\infty dy\,
\Omega_{\text{r}}(-2x+y+z)\,B_{\text{r}}(x,z)=0,\qquad y>0,$$
with the input data
$$\Omega_{\text{r}}(y):=\displaystyle\frac{1}{2\pi}\int_{-\infty}^\infty
dk\,L(k)\,e^{iky}+\displaystyle\sum_{j=1}^N c_{\text{r}j}^2\,
e^{-\tau_j y}.$$

\head 5. Krein's Spectral Shift Function \endhead

In this section, we indicate the construction of Krein's spectral
shift function $\xi_{\beta\alpha}(k)$
for $k\in\bold R^+\cup\bold I^+$
with $0<\beta<\alpha\le \pi,$ and we also show how to
extract $\alpha,$ $\beta,$ $F_\alpha,$ $F_\beta,$ and
$V$ from $\xi_{\beta\alpha}.$

Krein's spectral shift function $\xi_{\beta\alpha}(k)$ associated with
$H_\alpha$ and $H_\beta$
can be defined in various
ways [5,13-15]. We find it convenient to introduce $\xi_{\beta\alpha}$
by relating it to the phase of $F_\alpha(k)/F_\beta(k)$ for
$k\in\bold R\cup\bold I^+.$ Let us note that $\xi_{\beta\alpha}$
in [6]
is considered when
$0\le \beta<\alpha<\pi$, but studying it for
$0<\beta<\alpha\le \pi$ does not present any inconvenience
because we can choose $\xi_{0\theta}(k)=\xi_{\theta\pi}(k)$
for $k\in\bold R^+\cup\bold I^+$ with
$\theta\in(0,\pi),$ as done in our paper.

Let
$$e^{\pi i \xi_{\beta\alpha}(k)}:=\displaystyle\frac{Z_{\beta\alpha}(k)}
{|Z_{\beta\alpha}(k)|},\qquad k\in \bold R^+\cup\bold I^+,\tag 5.1$$
with the normalization
$$\xi_{\beta\alpha}(+\infty)=\cases 0,\qquad \alpha\in(0,\pi),\\
1/2,\qquad \alpha=\pi,\endcases\tag 5.2$$
and the range of $\xi_{\beta\alpha}(k)$ on $\bold I^+$ restricted
to the interval $[0,1],$
where we have defined
$$Z_{\beta\alpha}(k):=\cases \displaystyle\frac{F_\alpha(k)}{F_\beta(k)},\qquad 0<\beta<\alpha<\pi,\\
\displaystyle\frac{i\,F_\beta(k)}{F_\pi(k)}
,\qquad 0<\beta<\alpha=\pi.\endcases\tag 5.3$$
As seen from (2.5) and (5.1)-(5.3), we can express $\xi_{\beta\alpha}(k)$
for $k\in\bold R^+$
in terms of the phase shifts $\phi_\alpha$ and $\phi_\beta$ as
$$\xi_{\beta\alpha}(k)=\cases \displaystyle\frac{1}{\pi}\left[
\phi_\beta(k)-\phi_\alpha(k)\right],\qquad 0<\beta<\alpha<\pi,\\
\displaystyle\frac{1}{2}+\displaystyle\frac{1}{\pi}\left[
\phi_\pi(k)-\phi_\beta(k)\right],\qquad 0<\beta<\alpha=\pi.\endcases\tag 5.4$$
With the help of (2.6) and (5.4) we see
that
$\xi_{\beta\alpha}(k)$ can be extended from $k\in\bold R^+$ to
$k\in\bold R$ oddly, i.e.
$$\xi_{\beta\alpha}(-k)=
-\xi_{\beta\alpha}(k),\qquad 0<\beta<\alpha\le\pi,
\quad k\in\bold R.\tag 5.5$$

Using (3.18) of [12], for $k\in\bold R$ we get
$$\text{Im}\left[Z_{\beta\alpha}(k)\right]=\cases
\displaystyle\frac{k\,h_{\beta\alpha}}{|F_\beta(k)|^2},\qquad 0<\beta<\alpha< \pi,\\
\displaystyle\frac{k}{|F_\pi(k)|^2},
\qquad 0<\beta<\alpha=\pi.\endcases\tag 5.6$$
Since $F_\alpha(k)$ is nonzero
for $k\in\bold R^+,$
(5.6) implies that $\text{Im}\left[Z_{\beta\alpha}(k)\right]>0$ for
$k\in\bold R^+.$ Thus, from (5.1) we can conclude that $\xi_{\beta\alpha}(k)\in(0,1)$
when $k\in\bold R^+.$
Furthermore, using (2.8) and (5.4) we get
$$\xi_{\beta\alpha}(0^+)=\cases
N_\beta-N_\alpha+\displaystyle\frac{d_\beta-d_\alpha}{2},\qquad 0<\beta<\alpha<\pi,\\
N_\pi-N_\beta+\displaystyle\frac{d_\pi-d_\beta}{2}
,\qquad 0<\beta<\alpha=\pi.\endcases$$

Having introduced $\xi_{\beta\alpha}(k)$ for $k\in\bold R\cup\bold I^+,$
let us now analyze the problem of recovering
$\alpha,$ $\beta,$ $F_\alpha,$ $F_\beta,$ and
$V$ from $\xi_{\beta\alpha}.$
First, given $\xi_{\beta\alpha}(k)$ for $k\in\bold R^+\cup\bold I^+,$ from (5.2)
we determine whether $\alpha\in(0,\pi)$ or $\alpha=\pi.$

Next, we can recover $N_\alpha,$ $N_\beta,$ $\{\kappa_{\alpha j}\}_{j=1}^{N_\alpha},$
and $\{\kappa_{\beta j} \}_{j=1}^{N_\beta}$ by using the values of
$\xi_{\beta\alpha}(k)$ for $k\in\bold I^+.$ In fact, with the help of (2.1), (2.3),
and (5.1)-(5.3),
we see that $Z_{\beta\alpha}(k)$ is real valued for $k\in\bold I^+$ and
that $\xi_{\beta\alpha}(k)$ on $\bold I^+$
is equal to either $0$ or $1,$
with jump discontinuities at
$k=i\kappa_{\alpha j}$ and $k=i\kappa_{\beta j}.$
In other words, in consonant with the interlacing properties given in (3.1) and (3.2),
as a result of the simplicity of zeros of
$F_\alpha$ and $F_\beta$ on $\bold I^+,$
we have, when $N_\alpha=N_\beta$ and $0<\beta<\alpha<\pi$
$$\xi_{\beta\alpha}(i\omega)=\cases 0,\qquad \omega\in (0,\kappa_{\alpha 1})
\cup (\kappa_{\beta N_\beta},+\infty)
\cup_{j=2}^{N_\alpha} (\kappa_{\beta (j-1)}, \kappa_{\alpha j}),\\
1,\qquad \omega\in \cup_{j=1}^{N_\alpha} (\kappa_{\alpha j}, \kappa_{\beta j}),\endcases$$
and we have, when $N_\alpha=N_\beta-1$ and $0<\beta<\alpha<\pi$
$$\xi_{\beta\alpha}(i\omega)=\cases 0,\qquad \omega\in
(\kappa_{\beta N_\beta},+\infty)
\cup_{j=1}^{N_\alpha} (\kappa_{\beta j}, \kappa_{\alpha j}),\\
1,\qquad \omega\in (0,\kappa_{\beta 1})
\cup_{j=1}^{N_\alpha} (\kappa_{\alpha j}, \kappa_{\beta (j+1)})
.\endcases$$
On the other hand, we have,
when $0<\beta<\alpha=\pi$ and $N_\beta=N_\pi$
$$\xi_{\beta\pi}(i\omega)=\cases 1,\qquad \omega\in (0,\kappa_{\pi 1})
\cup (\kappa_{\beta N_\beta},+\infty)
\cup_{j=2}^{N_\pi} (\kappa_{\beta (j-1)}, \kappa_{\pi j}),\\
0,\qquad \omega\in \cup_{j=1}^{N_\pi} (\kappa_{\pi j}, \kappa_{\beta j}),\endcases$$
and we have, when $0<\beta<\alpha=\pi$ and $N_\beta=N_\pi+1$
$$\xi_{\beta\pi}(i\omega)=\cases 1,\qquad \omega\in
(\kappa_{\beta N_\beta},+\infty)
\cup_{j=1}^{N_\pi} (\kappa_{\beta j}, \kappa_{\pi j}),\\
0,\qquad \omega\in (0,\kappa_{\beta 1})
\cup_{j=1}^{N_\pi} (\kappa_{\pi j}, \kappa_{\beta (j+1)})
.\endcases\tag 5.7$$
Thus, we are able to recover $N_\alpha,$ $N_\beta,$ $\{\kappa_{\alpha j}\}_{j=1}^{N_\alpha},$
and $\{\kappa_{\beta j} \}_{j=1}^{N_\beta}$ by analyzing the
location of the jumps of $\xi_{\beta\alpha}(k)$ for $k\in\bold I^+.$

In order to continue with the recovery,
let us define the `reduced' quantities identified with the
superscript $[0]$ as follows:
$$F_{\alpha}^{[0]}(k):=F_{\alpha}(k)
\displaystyle\prod_{j=1}^{N_\alpha}\displaystyle\frac{k+i\kappa_{\alpha j}}
{k-i\kappa_{\alpha j}},\qquad \alpha\in (0,\pi].$$
Note that $|F_{\alpha}^{[0]}(k)|=|F_{\alpha}(k)|$ for $k\in\bold R,$ and hence
as seen from (5.3) it is natural to let
$$Z_{\beta\alpha}^{[0]}(k):=\cases Z_{\beta\alpha}(k)
\displaystyle\prod_{j=1}^{N_\alpha}\displaystyle\frac{k+i\kappa_{\alpha j}}
{k-i\kappa_{\alpha j}}
\displaystyle\prod_{p=1}^{N_\beta}\displaystyle\frac{k-i\kappa_{\beta p}}
{k+i\kappa_{\beta p}},\qquad 0<\beta<\alpha< \pi,\\
Z_{\beta\pi}(k)
\displaystyle\prod_{j=1}^{N_\pi}\displaystyle\frac{k-i\kappa_{\pi j}}
{k+i\kappa_{\pi j}}
\displaystyle\prod_{p=1}^{N_\beta}\displaystyle\frac{k+i\kappa_{\beta p}}
{k-i\kappa_{\beta p}},
\qquad 0<\beta<\alpha=\pi,\endcases
\tag 5.8$$
so that $Z_{\beta\alpha}^{[0]}$ has no zeros or poles in
$\overline{\bold C^+}\setminus\{0\}.$

Let $\Cal W^{[0]}$ denote the class of functions $Y(k)$ that are
analytic in $\bold C^+$ and continuous in
$\overline{\bold C^+}\setminus\{0\}$ satisfying
$Y(-k)=Y(k)^\ast$ for $k\in\bold R$ and
$1+O(1/k)$ as $k\to\infty$ in $\overline{\bold C^+},$ and that
$Y$ is either continuous at $k=0$ or has a finite-order zero or pole
at $k=0.$ Let
$\Cal W$ denote the extended class of functions that differ
 from $Y(k)$ in
$\Cal W^{[0]}$ by a finite number of
multiplicative factors of
the form $\displaystyle\frac{k+ia_j}
{k+ib_j}$ with real constants $a_j$
and $b_j.$
For such functions $Y(k)$ in $\Cal W,$ we let
$\log \left(Y(k)\right)$ denote the branch of
the logarithm normalized as $\text{Im}\left[\log \left(Y(\infty)\right)\right]=0.$

\proclaim {Theorem 5.1}
The quantities $Z_{\beta\alpha}^{[0]}(k)$ for $0<\beta<\alpha< \pi$
and $Z_{\beta\pi}^{[0]}(k)/[ik]$ for $0<\beta<\pi$ each belong to
$\Cal W^{[0]},$ and for
$k\in\overline{\bold C^+}$ we have
$$Z_{\beta\alpha}^{[0]}(k)=\exp\left(\displaystyle\frac{1}{\pi}
\int_{-\infty}^\infty dt\,
\displaystyle\frac{\text{\rm Im}\left[\log \left(Z_{\beta\alpha}^{[0]}(t)\right)\right]}
{t-k-i0^+}\right),\qquad 0<\beta<\alpha< \pi,\tag 5.9$$
$$Z_{\beta\pi}^{[0]}(k)=ik\cdot\exp\left(\displaystyle\frac{1}{\pi}
\int_{-\infty}^\infty dt\,
\displaystyle\frac{\text{\rm Im}\left[\log \left(Z_{\beta\pi}^{[0]}(t)/[it]\right)\right]}
{t-k-i0^+}\right),\qquad 0<\beta<\pi.\tag 5.10$$
\endproclaim

\demo{Proof}
For any $\alpha\in(0,\pi]$ it is known [2,10-12] that
$F_\alpha(k)$ is analytic in $\bold C^+$ and continuous in
$\overline{\bold C^+},$ it is nonzero in $\overline{\bold C^+}$
except for the simple zeros at $k=i\kappa_{\alpha j}$ with
$j=1,\dots,N_\alpha$ and perhaps a simple zero at $k=0.$
By using (3.12) and (3.13) of [12],
as $k\to\infty$ in $\overline{\bold C^+}$ we obtain
$$Z_{\beta\alpha}(k)
=\cases
1+\displaystyle\frac{i\,h_{\beta \alpha}}{k}
-\displaystyle\frac{h_{\beta\alpha}\,\cot\beta}{k^2}+o(1/k^2),
\qquad 0<\beta<\alpha< \pi,\\
ik+\cot\beta
+o(1),\qquad 0<\beta<\alpha=\pi,\endcases\tag 5.11$$
and hence from (5.8) and (5.11),
as $k\to\infty$ in $\overline{\bold C^+},$ we get
$$Z_{\beta\alpha}^{[0]}(k)
=
1-\displaystyle\frac{i}{k}\left[
h_{\beta \alpha}+2\sum_{j=1}^{N_\pi}\kappa_{\pi j}-2
\sum_{j=1}^{N_\beta}\kappa_{\beta j}\right]+O(1/k^2),
\qquad 0<\beta<\alpha< \pi,$$
$$\displaystyle\frac{Z_{\beta\pi}^{[0]}(k)}{ik}=
1+\displaystyle\frac{1}{ik}\left[
\cot\beta+2\sum_{j=1}^{N_\pi}\kappa_{\pi j}-2
\sum_{j=1}^{N_\beta}\kappa_{\beta j}\right]+o(1/k),
\qquad 0<\beta<\pi.$$
Using also (2.4), (5.3), and (5.8),
it follows that $Z_{\beta\alpha}^{[0]}(k)$ for $0<\beta<\alpha< \pi$
and $Z_{\beta\pi}^{[0]}(k)/[ik]$ for $0<\beta<\pi$ each belong to
$\Cal W^{[0]}.$
Moreover, the logarithms of both these quantities
are analytic in $\bold C^+$ and continuous in
$\overline{\bold C^+}\setminus\{0\},$ have at most an
integrable singularity at $k=0,$ and
are $O(1/k)$ as
$k\to\infty$ in $\overline{\bold C^+}.$
Thus, the Schwarz integral formula can be used to obtain
$$\log \left(Z_{\beta\alpha}^{[0]}(k)\right)=\displaystyle\frac{1}{\pi}
\int_{-\infty}^\infty dt\,
\displaystyle\frac{\text{Im}\left[\log \left(Z_{\beta\alpha}^{[0]}(t)\right)\right]}
{t-k-i0^+},\qquad k\in\overline{\bold C^+},$$
$$\log \left(\displaystyle\frac{Z_{\beta\pi}^{[0]}(k)}{ik}\right)=\displaystyle\frac{1}{\pi}
\int_{-\infty}^\infty dt\,
\displaystyle\frac{\text{Im}\left[\log \left(Z_{\beta\pi}^{[0]}(t)/[it]\right)\right]}
{t-k-i0^+},\qquad k\in\overline{\bold C^+},$$
which give us (5.9) and (5.10), respectively.
\quad\qed
\enddemo

 From (5.8) we get $|Z_{\beta\alpha}^{[0]}(k)|=|Z_{\beta\alpha}(k)|$
for $k\in\bold R,$ and hence
a comparison with (5.1) and (5.2) leads us to let
$$e^{\pi i \xi_{\beta\alpha}^{[0]}(k)}:=\displaystyle\frac{Z_{\beta\alpha}^{[0]}(k)}
{|Z_{\beta\alpha}^{[0]}(k)|},\qquad k\in \bold R^+\cup\bold I^+,\tag 5.12$$
with the normalization
$$\xi_{\beta\alpha}^{[0]}(+\infty)=\cases 0,\qquad \alpha\in(0,\pi),\\
1/2,\qquad \alpha=\pi,\endcases$$
so that for $0<\beta<\alpha< \pi$ we have
$$\xi_{\beta\alpha}^{[0]}(k)=\xi_{\beta\alpha}(k)+
\displaystyle\frac{1}{\pi i}
\log\left(\displaystyle\prod_{j=1}^{N_\alpha}\displaystyle\frac{k+i\kappa_{\alpha j}}
{k-i\kappa_{\alpha j}}
\displaystyle\prod_{p=1}^{N_\beta}\displaystyle\frac{k-i\kappa_{\beta p}}
{k+i\kappa_{\beta p}}\right),\tag 5.13$$
and for $0<\beta<\pi$ we get
$$\xi_{\beta\pi}^{[0]}(k)=\xi_{\beta\pi}(k)+
\displaystyle\frac{1}{\pi i}
\log\left(\displaystyle\prod_{j=1}^{N_\pi}\displaystyle\frac{k-i\kappa_{\pi j}}
{k+i\kappa_{\pi j}}
\displaystyle\prod_{p=1}^{N_\beta}\displaystyle\frac{k+i\kappa_{\beta p}}
{k-i\kappa_{\beta p}}\right).\tag 5.14$$

Using (5.5), (5.13), and (5.14) we can extend $\xi_{\beta\alpha}^{[0]}(k)$
oddly from $k\in\bold R^+$ to $k\in\bold R,$ i.e.
$$\xi_{\beta\alpha}^{[0]}(-k)=
-\xi_{\beta\alpha}^{[0]}(k),\qquad 0<\beta<\alpha\le \pi,
\quad k\in\bold R.\tag 5.15$$
 From (5.12) and (5.15), for $k\in\bold R$ we get
$$\xi_{\beta\alpha}^{[0]}(k)=\cases\displaystyle\frac{1}{\pi}\,\text{Im}\left[
\log \left(Z_{\beta\alpha}^{[0]}(k)\right)\right],\qquad 0<\beta<\alpha< \pi,\\
\displaystyle\frac{1}{\pi}\,\text{Im}\left[
\log \left(\displaystyle\frac{Z_{\beta\pi}^{[0]}(k)}{ik}
\right)\right]+\displaystyle\frac12\,\text{sign}(k),\qquad
0<\beta<\alpha=\pi,\endcases\tag 5.16$$
and similarly, for $k\in\bold R$ we have
$$\xi_{\beta\alpha}(k)=\cases\displaystyle\frac{1}{\pi}\,\text{Im}\left[
\log \left(Z_{\beta\alpha}(k)\right)\right],\qquad 0<\beta<\alpha< \pi,\\
\displaystyle\frac{1}{\pi}\,\text{Im}\left[
\log \left(\displaystyle\frac{Z_{\beta\pi}(k)}{ik}\right)\right]+\displaystyle\frac12\,\text{sign}(k),\qquad
0<\beta<\alpha=\pi.\endcases\tag 5.17$$
Since $Z_{\beta\alpha}^{[0]}(k)>0$ for $k\in\bold I^+,$
we have
$$\xi_{\beta\alpha}^{[0]}(k)=0,\qquad 0<\beta<\alpha\le\pi,
\quad k\in\bold I^+,\tag 5.18$$
and hence, from (5.13) and (5.14), for $k\in\bold I^+$ we get
$$\xi_{\beta\alpha}(k)=\cases
\displaystyle\frac{1}{\pi i}
\log\left(\displaystyle\prod_{j=1}^{N_\alpha}\displaystyle\frac{k-i\kappa_{\alpha j}}
{k+i\kappa_{\alpha j}}
\displaystyle\prod_{p=1}^{N_\beta}\displaystyle\frac{k+i\kappa_{\beta p}}
{k-i\kappa_{\beta p}}\right),\qquad 0<\beta<\alpha< \pi,\\
\displaystyle\frac{1}{\pi i}
\log\left(\displaystyle\prod_{j=1}^{N_\pi}\displaystyle\frac{k+i\kappa_{\pi j}}
{k-i\kappa_{\pi j}}
\displaystyle\prod_{p=1}^{N_\beta}\displaystyle\frac{k-i\kappa_{\beta p}}
{k+i\kappa_{\beta p}}\right),\qquad
0<\beta<\alpha=\pi.\endcases\tag 5.19$$

Using (5.16) in (5.9) and (5.10), we obtain the following.
\proclaim {Corollary 5.3} For $0<\beta<\alpha\le \pi,$
the quantity $Z_{\beta\alpha}^{[0]}(k)$ for $k\in\overline{\bold C^+}$ can be obtained
 from $\xi_{\beta\alpha}^{[0]}(k)$ given for $k\in\bold R^+$ via (5.15) and
$$Z_{\beta\alpha}^{[0]}(k)=\cases
\exp\left(
\displaystyle\int_{-\infty}^\infty dt\,
\displaystyle\frac{\xi_{\beta\alpha}^{[0]}(t)}
{t-k-i0^+}\right),\qquad 0<\beta<\alpha< \pi,\\
ik\cdot\exp\left(
\displaystyle\int_{-\infty}^\infty dt\,
\displaystyle\frac{\xi_{\beta\pi}^{[0]}(t)-(1/2)\,\text{\rm sign}(t)
}
{t-k-i0^+}\right),\qquad 0<\beta<\alpha=\pi.\endcases$$
\endproclaim

Now let us continue with
the recovery of $\alpha$ and $\beta$ from $\xi_{\beta\alpha}.$
We have already constructed
$N_\alpha,$ $N_\beta,$ $\{\kappa_{\alpha j}\}_{j=1}^{N_\alpha},$
and $\{\kappa_{\beta j} \}_{j=1}^{N_\beta}.$
Next, with the help of (5.13), (5.14), and (5.15) we obtain
$\xi_{\beta\alpha}^{[0]}(k)$ for $k\in\bold R.$ Then, using
Corollary~5.3 we get $Z_{\beta\alpha}^{[0]}(k)$ and via
(5.8) we obtain $Z_{\beta\alpha}(k)$ for $k\in\overline{\bold C^+}.$
Having $Z_{\beta\alpha}(k)$ in hand, we can recover $\alpha$ and $\beta$
by using (3.11) and (5.11).

Next, we continue with the construction of $F_\alpha$ and
$F_\beta.$ When $\alpha\ne\pi,$ we proceed as follows.
Having $h_{\beta\alpha}$ and $Z_{\beta\alpha}(k)$ for $k\in\bold R,$
via (5.6) we construct $|F_\beta(k)|^2$ as
$$|F_\beta(k)|^2=
\displaystyle\frac{k\,h_{\beta\alpha}}{\text{Im}\left[Z_{\beta\alpha}(k)\right]},
\qquad k\in\bold R.$$
Knowing $|F_\beta(k)|$ for $k\in\bold R,$ we
can then construct $F_\beta$ via (cf. (3.6) of [12])
$$F_\beta(k)=k\left(\displaystyle\prod_{j=1}^{N_\beta}\displaystyle\frac{k-i\kappa_{\beta
j}}{k+i\kappa_{\beta j}}\right) \exp\left(\displaystyle\frac{-1}{\pi
i}\int_{-\infty}^\infty dt\, \displaystyle\frac{\log
|t/F_\beta(t)|}{t-k-i0^+}\right), \qquad k\in\overline{\bold C^+}.\tag 5.20$$
On the other hand, if $\alpha=\pi$ we proceed as follows. From (5.6)
we get $|F_\pi(k)|^2$ as
$$|F_\pi(k)|^2=
\displaystyle\frac{k}{\text{Im}\left[Z_{\beta\pi}(k)\right]},
\qquad k\in\bold R.
$$
Having $|F_\pi(k)|$ for $k\in\bold R$ at hand, we
can then construct $F_\pi$ via (cf. (3.7) of [12])
$$F_\pi(k)=\left(\displaystyle\prod_{j=1}^{N_\pi}
\displaystyle\frac{k-i\kappa_{\pi j}}{k+i\kappa_{\pi j}}\right)
\exp\left(\displaystyle\frac{1}{\pi i}\int_{-\infty}^\infty dt\,
\displaystyle\frac{\log |F_\pi(t)|}{t-k-i0^+}\right),\qquad k\in\overline{\bold C^+}.\tag 5.21$$

Finally, if $\alpha\ne \pi$ then via the first line of (5.3) we
can recover $F_\alpha(k)$ from the already constructed
quantities $F_\beta(k)$ and $Z_{\beta\alpha}(k)$ for
$k\in\overline{\bold C^+}.$ Similarly,
if $\alpha=\pi$ then via the second line of (5.3) we
can recover $F_\beta(k)$ from the already constructed
quantities $F_\pi(k)$ and $Z_{\beta\pi}(k)$ for
$k\in\overline{\bold C^+}.$
A comparison with (3.3)-(3.10) reveals that
we have now constructed
the data sets $\Cal D_1,\Cal D_3,\Cal D_5,\Cal D_7$ when
$\alpha\in(0,\pi)$ and $\Cal D_2,\Cal D_4,\Cal D_6,\Cal D_8$ when
$\alpha= \pi.$ Finally, we can recover $V$ via
one of the methods described in Section~4.

Let us also note that, by using any one of the eight data
sets $\Cal D_j$ given in (3.3)-(3.10), we can construct
the quantities $F_\alpha(k)$ and $F_\beta(k)$ for $k\in\overline{\bold C^+}$ as outlined in Section~3,
then obtain $Z_{\beta\alpha}$ for $k\in\overline{\bold C^+}$ given
in (5.3), and also recover
$\xi_{\beta\alpha}(k)$ for $k\in\bold R^+\cup\bold I^+$ via
(5.17) and (5.19).

\head 6. Examples \endhead

In this section we present two examples to illustrate
the recovery of the potential and the boundary
conditions from the data set $\Cal D_3$ given in (3.5)
and also from Krein's spectral shift function $\xi_{\beta\alpha}.$

\example{Example 6.1} In our first example, assume that we are given $\Cal D_3$
with $h_{\beta\alpha}=5,$ $N_\alpha=1,$ $N_\beta=2,$ $\kappa_{\alpha 1}=2,$
$\kappa_{\beta 2}=4,$ and $|F_\alpha(k)|^2=k^2+4$ for $k\in\bold R,$
and we are interested in constructing all the relevant quantities
such as $\cot\alpha,$ $\cot\beta,$ $F_\alpha,$ $F_\beta,$ and $V.$
By using the method outlined in Section~3, with the details
given in the proof of Theorem~2.3 of [12], we get
$$\kappa_{\beta 1}=1,\quad \cot\alpha=-8/5,\quad \cot\beta=17/5,$$
$$F_\alpha(k)=k-2i,\quad F_\beta(k)=\displaystyle\frac{(k-i)(k-4i)}{k+2i}.$$
As indicated in Sections~3-5, we can then obtain other relevant
quantities such as
$$f(k,0)=\displaystyle\frac{5k-8i}{5(k+2i)},\quad
f'(k,0)=\displaystyle\frac{25ik^2+40k+36i}{25(k+2i)},
$$
the quantities used in the Gel'fand-Levitan and Marchenko methods, e.g.
$$g_{\alpha 1}=\sqrt{\displaystyle\frac 25},\quad
m_{\alpha 1}=\sqrt{40},\quad
S_\alpha(k)=\displaystyle\frac{k+2i}{k-2i},$$
the quantities used in the Faddeev-Marchenko method, e.g.
$$N=1,\quad \tau_1=\displaystyle\frac{(\sqrt{34}+4)i}{5},
\quad c_{\text{r}1}=\displaystyle\frac{3}{\sqrt{5\sqrt{34}}},
\quad L(k)=\displaystyle\frac{-18}{25k^2-40ik+18},$$
the quantities relevant to Krein's spectral shift function, e.g.
$$Z_{\beta\alpha}^{[0]}(k)=\displaystyle\frac{(k+2i)^2}
{(k+i)(k+4i)},\quad
\xi_{\beta\alpha}^{[0]}(k)=\displaystyle\frac{1}{\pi}\,
\text{Im}\left[\log \left(\displaystyle\frac{(k+2i)^2}
{(k+i)(k+4i)}\right)\right],$$
$$Z_{\beta\alpha}(k)=\displaystyle\frac{(k+2i)(k-2i)}
{(k-i)(k-4i)},\quad
\xi_{\beta\alpha}(k)=\displaystyle\frac{1}{\pi}\,
\text{Im}\left[\log \left(\displaystyle\frac{(k+2i)(k-2i)}
{(k-i)(k-4i)}\right)\right],$$
and finally the potential and the Jost solution
$$V(x)=-\displaystyle\frac{288e^{4x}}
{\left(9+e^{4x}\right)^2},\quad
f(k,x)=e^{ikx}\left[1-\displaystyle\frac{36i}{(k+2i)\left(9+e^{4x}\right)}\right].$$
\endexample

\example{Example 6.2} As our second example, let us assume that we have as our data
Krein's spectral shift function given by
$$\xi_{\beta\alpha}(k)=\cases \displaystyle\frac12+
\displaystyle\frac{1}{\pi}\,
\text{Im}\left[\log \left(\displaystyle\frac{(k-i)(k+i)(k-3i)}
{k(k-2i)(k+2i)}\right)\right],\qquad k\in\bold R^+,\\
0,\qquad
k\in i(0,1)\cup i(2,3),\\
1,\qquad k\in i(1,2)\cup i(3,+\infty),\endcases
\tag 6.1$$
and we would like to construct all the relevant quantities
such as $\cot\alpha,$ $\cot\beta,$ $F_\alpha,$ $F_\beta,$ and $V.$
By letting $k\to+\infty$ in (6.1) we get $\xi_{\beta\alpha}(+\infty)=1/2,$
and hence we see from (5.2) that $\alpha=\pi.$ Next, we analyze
our $\xi_{\beta\alpha}(k)$ on $\bold I^+$ and see the jump discontinuities
at $k=i,2i,3i.$ A comparison with (5.7) indicates
that $N_\alpha=1,$ $N_\beta=2,$
$\kappa_{\alpha 1}=2,$ $\kappa_{\beta 1}=1,$ and
$\kappa_{\beta 2}=3.$ Next, using (5.14) and (5.18) we obtain
$$\xi_{\beta\alpha}^{[0]}(k)=\cases
\displaystyle\frac12
+\displaystyle\frac{1}{\pi}\,
\text{Im}\left[\log \left(\displaystyle\frac{(k+i)^2(k+3i)}
{k(k+2i)^2}\right)\right],\qquad k\in\bold R,\\
0,\qquad k\in\bold I^+,\endcases$$
and then, via Corollary~5.3, we get
$$Z_{\beta\alpha}^{[0]}(k)=\displaystyle\frac
{i(k+i)^2(k+3i)}{(k+2i)^2}.\tag 6.2$$
Next, using (6.2) in (5.8) we obtain
$$Z_{\beta\alpha}(k)=\displaystyle\frac
{i(k+i)(k-i)(k-3i)}{(k+2i)(k-2i)}.\tag 6.3$$
Using (6.3) in (5.11) we get $\cot\beta=3,$ and then
with the help of (5.20), (5.21), and then (5.3)
we obtain
$$F_\pi(k)=\displaystyle\frac
{k-2i}{k+i},\quad
F_\beta(k)=\displaystyle\frac
{(k-i)(k-3i)}{k+2i}.$$
Proceeding as in the first example, we
can then construct all the relevant quantities. For example, we have
$$S_\pi(k)=\displaystyle\frac{(k+i)(k+2i)}{(k-i)(k-2i)},\quad
m_{\pi 1}=4\sqrt{3},\quad g_{\pi 1}=\sqrt{3},$$
$$L(k)=\displaystyle\frac{-9i}{2k^3+5k+9i},\quad
N=1,\quad \tau_1=2.14444\overline{1}\,i,\quad
c_{{\text r}1}=0.63118\overline{2},$$
where the overline indicates a roundoff at the digit.
Finally, we obtain the potential and the Jost solution
$$V(x)=\displaystyle\frac{24e^{-2x}-480e^{-4x}+720e^{-6x}-480e^{-8x}+600e^{-10x}}
{\left(1-3e^{-2x}+15e^{-4x}-5e^{-6x}\right)^2},$$
$$
f(k,x)=e^{ikx}\left[1+\displaystyle\frac{\displaystyle\frac{6i(e^{-2x}-5e^{-6x})}{k+i}+
\displaystyle\frac{60i(-e^{-4x}+e^{-6x})}{k+2i}}
{1-3e^{-2x}+15e^{-4x}-5e^{-6x}}
\right].$$
\endexample

\refstyle{N}

\enddocument